\documentclass[aps,prd,groupedaddress,twocolumn]{revtex4}
\usepackage[pdfpagemode=FullScreen,bookmarks=true]{hyperref}
\usepackage[dvips]{graphicx}
\usepackage[latin1]{inputenc}
\begin{document}

\title{Light Higgs Boson Production in Two Higgs Doublets Models type III}
\author{Camilo Andr\'es Jim\'enez-Cruz}
\email{cajimenezc@unal.edu.co}

\author{Roberto  Martinez}
%\email{remartinezm@unal.edu.co}

\author{J.-Alexis Rodríguez López}
\email{jarodriguezl@unal.edu.co}
\affiliation{Departamento de Física - Universidad Nacional de Colombia - Sede Bogota, Colombia}

\hbadness 100000
\vbadness 100000

\date{\today}

\begin{abstract}
By using the Cheng, Sher and Yuan's anzats, we study the light Higgs Boson production associated with $b$ quark production at TEVATRON using the 2HDM type III.
 We compare the simulations with experimental results coming from TEVATRON, finding valid ranges for the $bb$ coupling.
By using these results, we calculate the cross section for the process $pp \to b\bar bh(b\bar b)$ for the LHC collider.
\end{abstract}

 %ToDo:
 %ToDo:

\maketitle

\section{Introduction}
Although a relatively light Higgs boson coming from the SM can support  the idea of the SM being true all the way to the Plank scale, most theorists consider such a possibility unlikely\cite{carena-2000}. 
A light Higgs boson is preferred by precision fits of the Standard Model (SM)\cite{collaboration-2007} and also theoretically required by many frameworks that can be effectively SM-like at low energies. The production of a Higgs boson in association with a heavy quark and antiquark pair (both $t\bar t$ and $b\bar b$) at hadron colliders will be sensitive to the Higgs-fermion couplings and can help discriminate between models.
One of such models is the so-called 2HDM, which includes a second scalar doublet with the same properties of the first one.\\

In this paper, by using the framework of the 2HDM-III and the CSY anzats (section \ref{s:2hdm}), we calculate the associated b-quark/Higgs boson production at the TEVATRON collider. By comparing the results obtained in the simulation with experimental data released by the D0 collaboration we find bounds over the $\lambda_{bb}$ parameter of the coupling (section \ref{s:associated}). The bounds found let us estimate the associated higgs/bottom production in the Large Hadron Collider (section \ref{s:results}). Finally, in section \ref{s:conc} we state our concluding remarks.

\section{\label{s:2hdm}Two Higgs Doublet Model (2HDM)}
The 2HDM includes a second scalar doublet, with each doublet acquiring a vacuum expectation value different from zero.\\

The scalar content of the model becomes:
\begin{equation}
\Phi_i=\left(\begin{array}{c}
 \phi_i^+\\
 \phi_i^0
\end{array}\right), \ \ \ \ 
\left<\Phi_i\right>=\left(\begin{array}{c}
 0\\
 \frac{v_i}{\sqrt2}
\end{array}\right), \ \ \ \
i=1,2 
\end{equation}
In this way, the scalar spectrum of mass eigenvalues contains two CP-even neutral Higgs bosons $(h_0,H_0)$ coming from the mixing of the real part of the neutral components of both doublets with a mixing angle $\alpha$; two charged Higgs bosons ($H^{\pm})$, which mix with the would-be Goldstone bosons ($G_W^{\pm}$) through the mixing angle $\tan\beta =v_2/v_1$ and one CP-odd Higgs ($A_0$), which mixes  with the neutral would-be Goldstone.
\subsection{The Yukawa Lagrangian}
\begin{table}[b]
\begin{center}
\begin{tabular}{cc} \hline Parameter&Range\\\hline\hline
$\xi_{\mu\tau}^2$                       &  [7.62$\times 10^{-4}:4.44\times10^{-2}$]\\
$\xi_{\tau\tau}$                        &  [-1.8$\times 10^{-2}:2,2\times10^{-2}$]\\
$\xi_{\mu\mu}$                          &  [-0.12:0.12]\\
$\xi_{\mu e}$                           &  [-0.39:0.39]\\
$\lambda_{bb}$                          &  [-100:100]\\
$\lambda_{tt}$                          &  [-$\sqrt 8: \sqrt 8$]\\\hline
\end{tabular}
\end{center}
\caption{\label{t:bounds} Experimental constraints over the $\xi$ and $\lambda$ matrices}
\end{table}
The most general Lagrangian that can be written in this kind of models includes interactions between all the fermions and both doublets:
\begin{eqnarray}
-\mathcal L_Y&=&
\eta_{ij}^{U,0}\bar Q_{iL}^0\tilde\Phi_1 U_{jR}^0+\nonumber
\eta_{ij}^{D,0}\bar Q_{iL}^0\tilde\Phi_1 D_{jR}^0\\
&+&\xi_{ij}^{U,0}\bar Q_{iL}^0\tilde\Phi_2 U_{jR}^0+
\xi_{ij}^{D,0}\bar Q_{iL}^0\tilde\Phi_2 D_{jR}^0\nonumber\\
&+& \mbox{l.s.\ +\ h.c}
\end{eqnarray}
The strong suppression of FCNC at tree level makes customary to impose discrete symmetries over the doublets.\\
Those symmetries end in one of three types of models\cite{diaz-2002}.
\begin{itemize}
 \item {\bf Type I}\\ 
 $\Phi_1$ is the responsible of giving mass to the matter, while $\Phi_2$ decouples totally from them.
 \item {\bf Type II}\\ $\Phi_1$ couples to the $up$ sector while $\Phi_2$ couples to the down sector. This is the case of the MSSM.
 \item {\bf Type III}\\ Both doublets couple to both sectors. 
\end{itemize}
\subsection{2HDM type III}
As there are two  non-diagonal $3\times3$ matrices in the Yukawa lagrangian,  and the suffix 0 means that these fermion states are not mass eigenstates. It is clear that
the mass terms for the matter will depend on two Yukawa coupling
matrices. The rotation of the quarks and leptons allow us to diagonalize one of the matrices
but in general not both simultaneously, then one Yukawa coupling remains non-diagonal,
leading to the FCNC at tree level.
\subsection{Cheng, Sher and Yuan's Anzats (CSY)}
In this work, we use the CSY parameterization of the Yukawa's couplings\cite{sher-1998}. This anzats is based on the SM $\phi f\bar f$ couplings and states that 
$$\xi^{ij}\equiv\frac{\sqrt{m_im_j}}{v}\lambda_{ij}$$ 
This is an ansatz for the Yukawa texture matrices looking for a phenomenological similarity
with SM couplings. Here,  $v$ has been defined as $v\equiv\sqrt{v_1^2+v_2^2}$.\\
 
Some restrictions over the $\lambda_{ij}$ and the $\xi_{ij}$  parameter sets have been found\cite{martinez-2005-72} and can be seen in Table \ref{t:bounds}.
The $\lambda_{bb}$ parameter can be costrained by the coupling of the $b$ quark to the scalar sector.

\section{\label{s:associated}Associated $h_0b\bar b$ production}
\begin{figure}[t]
\begin{center}
 \includegraphics[width=.4\linewidth]{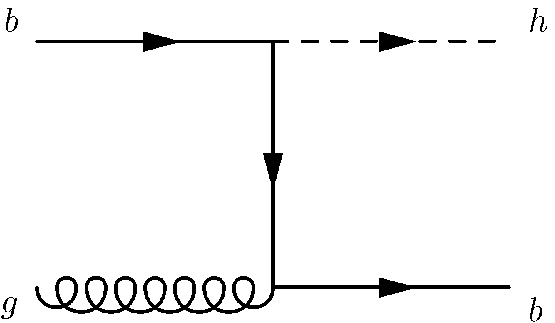}
 \includegraphics[width=.4\linewidth]{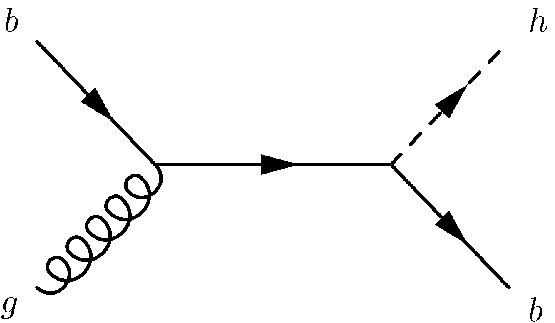}
\end{center}
\begin{center}
 \includegraphics[width=.4\linewidth]{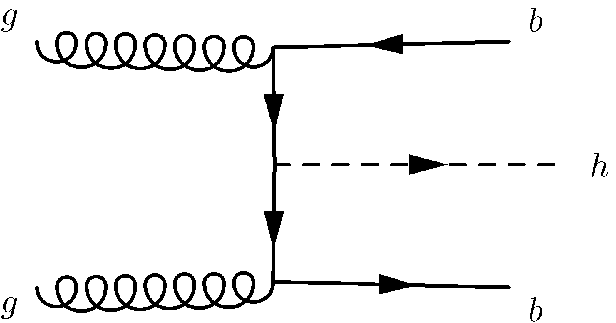}
 \includegraphics[width=.4\linewidth]{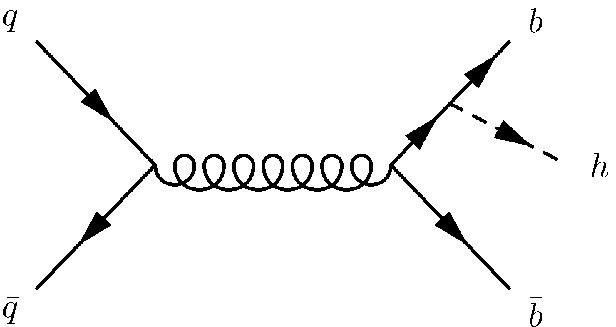}
\end{center}
\caption{\label{f:proc} Processes involved in the associated $\phi/b$ production}
\end{figure}
In the SM, the scalar sector couples to the fermions with strength proportional to the mass of the fermion. In this way, the Yukawa coupling to bottom quarks ($m_b\sim 5$GeV) leads to small cross sections in hadron colliders. However, processes involving $b$ quarks and light Higgs bosons (Figure \ref{f:proc}) can be strongly enhanced in some 2HDM scenaries. This associated production has been extensivelly studied for 2HDM type II\cite{a1,a2,a3,a4},
 being the Minimal Supersymetric Standar Model a favorite framework for calculation. However, 2HDM type III models' cross sections can be enhanced in a similar way.\\

The predominant modes are
\begin{equation}qq,gg\to b\bar b h_0(b\bar b),\mbox{ and }gb\to bh(b\bar b)\end{equation}
so the interesting events would have at least four jets in the final state where two of these jets should come from a Higgs resonance. The signal-to-background ratio can be enhanced by requiring that the four jets (or at least three) of the highest transverse energies are tagged as b-jets. This mode could be of great interest when the mass of the scalar boson is greater than 150MeV, where the sensitivity to the $h\to \tau\tau$ channel disappears\cite{atlas}.\\

Specifically, in the frame of the 2HDM-III and by using the CSY anzats, the coupling between the lightest Higgs boson and the $b$ quarks is given by
\begin{equation}A(h_0b\bar b) = h_0\bar b \left(
\frac{\cos\alpha+\sin\alpha\tan\beta}{\sqrt2}\xi^D
-\frac{\sin\alpha}{v}m_b
\right)b
\end{equation}
where $\alpha$ and $\beta$ are the angles defined before.\\

In terms of the {\it fundamental parameterization}\cite{diaz-2002}, the coupling becomes
\begin{equation}A(h_0b\bar b) = h_0\bar b \left(-\frac{\sin\alpha}{v}m_b+\frac{\cos\alpha}{\sqrt2}\frac{m_b}{v}\lambda_{bb}\right)b.\end{equation}
where we see explicitelly the high dependence on the $\lambda_{bb}$ parameter defined in the CSY anzats.

\begin{figure}[t]
\begin{center}
\includegraphics[width=\linewidth]{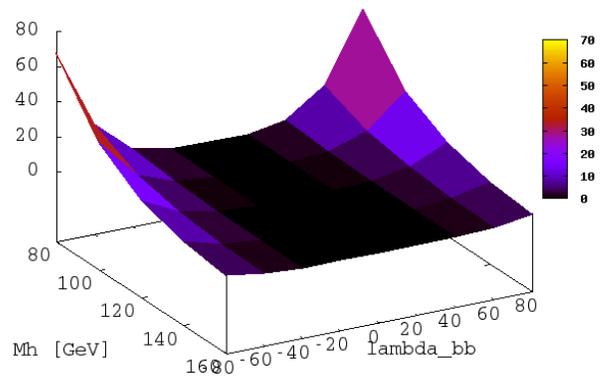}
\caption{\label{f:pp2bh-TEVATRON}Light Higgs boson production associated with two high $p_t$ b-quarks and decaying into a $b\bar b$ pair, for different values of $\lambda_{bb}$ and $M_h$, $\alpha=\sqrt2/2$}
\end{center}
\end{figure}
\section{Results\label{s:results}}
\begin{figure}
\begin{center}
\includegraphics[width=\linewidth]{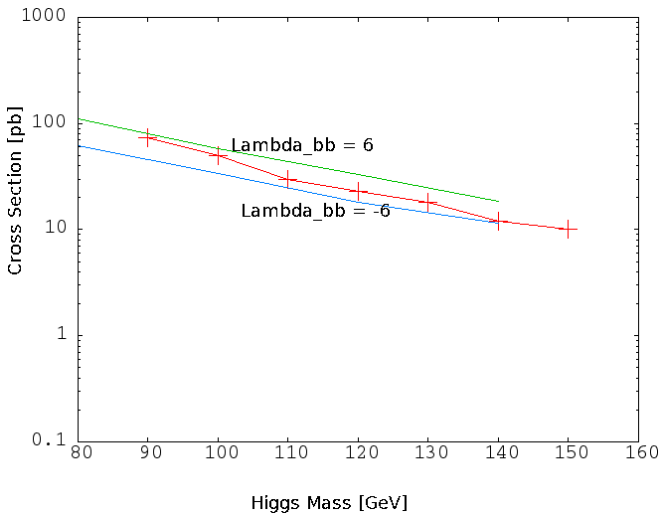}
\caption{\label{f:vstevatron}Plot matching experimental data\cite{collaboration-2005-95} comming from the D0 experiment (TEVATRON) and simulated data running $m_h$ and $\lambda_{bb}$.}
\end{center}
\end{figure}

\begin{figure}
\begin{center}
\includegraphics[width=\linewidth]{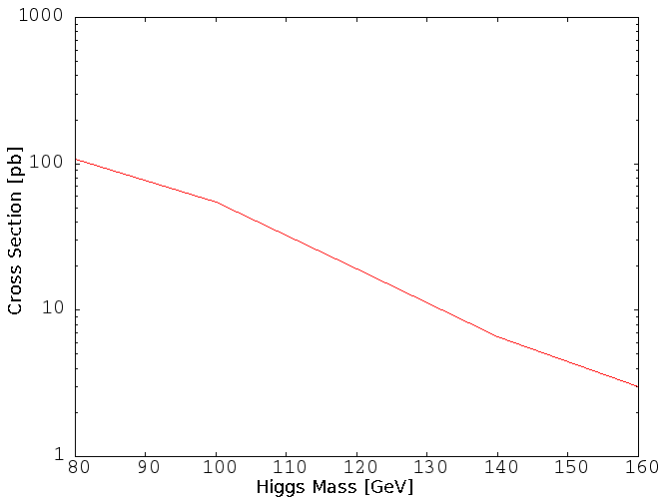}
\caption{\label{f:res}Estimated cross section for the associated bottom quark production for the LHC collider ($pp\to b\bar bh(b\bar b)$), here we used $\lambda_{bb}=6$}
\end{center}
\end{figure}
By using the CalcHEP package\cite{calchep} as the Monte Carlo event generator, we calculated the associated bottom-quark/Higgs production in the TEVATRON collider for different values of $\lambda_{bb}$ and $m_\phi$. The results can be seen in  Figure \ref{f:pp2bh-TEVATRON}. We used the CTEQ6M Parton Distribution Function for the proton and the antiproton, a momentum of 1980GeV at center of mass and a representative value of $\alpha=\pi/4$.\\

As in this parameter region the $h_0$  boson is nearly degenerate in mass with either the $h_0$ or the $A_0$ bosons, they cannot be  distinguished experimentally\cite{collaboration-2005-95} and the total cross section for the signal is assumed to be twice of the light boson\cite{carena-2006-45}.\\ 

The total cross section for the process is then\cite{carena-2006-45}:
\begin{equation}\sigma(p\bar p\to b\bar b h_0\to b\bar bb\bar b)\sim 2\times\sigma(p\bar p\to b\bar b\phi)\times BR(h_0\to b\bar b).\end{equation}

The results were matched against experimental results comming from the D0 experiment\cite{collaboration-2005-95} (Figure \ref{f:vstevatron}), showing a great agreement for small values of $\lambda_{bb}$.  In this way, we got the following bounds over the $\lambda_{bb}$ parameter, given $\alpha\sim\frac{\sqrt 2}2$:
\begin{equation}-6\le\lambda_{bb}\le6\end{equation}.

By using the bounds obtained, we can estimate the associated b-quark-higgs boson production rates for the LHC collider. The results are depicted in the figure \ref{f:res}.\\

\section{Concluding Remarks\label{s:conc}}
The coupling between the light neutral Higgs boson and the fermions is quite sensitive to the existence of a non-minimal scalar sector.\\
By using the Cheng, Sher and Yuang parameterization of the coupling between the $b$ quark and the scalar sector we calculated the total cross section for the process $p\bar p\to b\bar bh(b\bar b)$ at a center of mass energy of $1.98$TeV. We compared this result  with the reported data comming from  the D0 experiment at TEVATRON\cite{Mal2006}. In this way, we obtained that the  $\lambda_{bb}$ parameter of the anzats should be in the range $[-6,6]$ improving the actual bound provided by unitarity and included in the table \ref{t:bounds}.

With this results, we calculated the associated $b$ quark production of the light Higgs boson for the LHC.

\begin{acknowledgments}
R.M. acknowledges the finantial support of the Banco de la Republica de Colombia.
\end{acknowledgments}

\end{document}